# Creating temperature dependent Ni-MH battery models for low power mobile devices

*D. Szente-Varga, Gy. Horvath, M. Rencz*

(szvdom | horvath | rencz@eet.bme.hu)
Budapest University of Technology and Economics
Department of Electron Devices

## ABSTRACT

In this paper the methodology and the results of creating temperature dependent battery models for ambient intelligence applications is presented. First the measurement technology and the model generation process is presented in details, and then the characteristic features of the models are discussed.

## 1. INTRODUCTION

Mobile devices, like sensor networks and MEMS actuators use mobile power supplies to ensure energy for their operation. These are mostly batteries. The lifetime of the devices depends on the power consumption and on the quality and capacitance of the battery. Though the integrated circuits and their power consumption improves continually, their clock frequency also increases with the time, and the resultant power consumption seems not to vary, or slightly increase. On the other hand, the properties of batteries are developing much slower, necessitating the optimization of their usage on system level [5].

A possible solution to increase lifetime of mobile applications is to consider the battery-aware design and practice, which means that the application software and hardware has to shape the operation to the actual status of the supplying battery. To develop and design systems to handle this problem, battery models have to be available [3] [4].

This article presents the creation of temperature dependent Ni-MH battery models designed for systems optimization. In this paper the measuring methodology, measurement setup, results, observations and model generation principles of temperature dependent battery models are detailed.

## 2. MEASUREMENTS

Ni-MH batteries are the most frequently used mobile energy storage units today. For the measurement, we used four AAA size Ni-MH rechargeable batteries, two of them with a capacitance of 800mAh, and two of them of 850mAh. The batteries have been measured under constant resistant loads, at constant temperature values. Each measurement was started after the batteries had been fully charged, and the measurements were finished when the batteries became fully discharged by the loads, that is, their output voltage dropped under the preset threshold voltage. The same four batteries were used throughout the whole measurement sequence.

The batteries were measured at different temperature values, namely 1°C, 5°C, 10°C, 20°C, 30°C, 40°C, 50°C and 60°C with the value of 2.5Ω load resistance. This is 0.6C load current for the 800mAh batteries (which means that the load current was 0.6 times the capacity of the battery per hour, 800mA*0.6 = 480mA), and 0,56C for the 850mAh batteries.

The measurements with different loads were done at 20°C, and 30Ω, 15Ω, 7.5Ω, 3.75Ω, 2.5Ω, 1.88Ω and 1.5Ω resistors were used. Each measurement was performed at least twice. Discharging a 850mAh battery with 2.5Ω load resistance takes approximately 1 hour and 40 minutes, with 1.5Ω it is one hour. Discharging with 30Ω takes twenty hours. A charging period takes five hours, consequently these measurements are rather lengthy, and need to be computer controlled.

During one measurement four batteries were measured of the same value of load resistance and had the same temperature conditions.

The measurements were controlled and monitored by a personal computer. The four batteries were packaged and immersed into a temperature controlled water basin. The required water temperature was controlled by the T3Ster thermal transient tester's THERMOSTAT device [1] with a variance of 0.01°C. The terminals of the battery block from the THERMOSTAT were connected to a PCB, which contained the preset load resistors, the switching devices, and the control LEDs. This panel was in connection with the measuring equipment.

We used the T3Ster thermal transient tester [2] to measure the four batteries simultaneously. This device has got 8 channels to measure voltage with very high resolution: 200mV/1024. Counting with a little A/D converter error, we still get a sufficient resolution of about 1 mV to measure the batteries' output voltages during the discharge period.

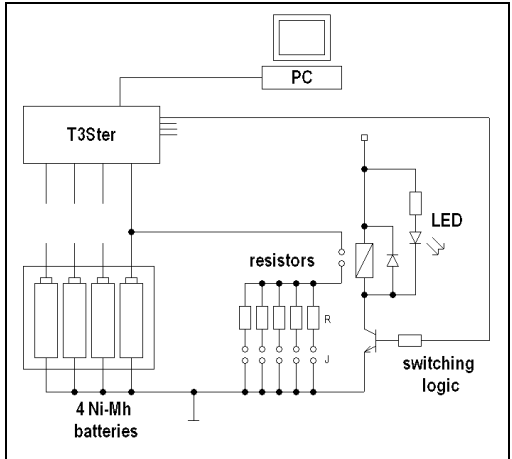

Fig. 1. Block-diagram of the measurement

The measured data were collected by the PC, which was directly connected to the tester. It also controlled the whole measurement processing and switched the loads off from a battery if it became fully discharged. The block diagram of the whole measurement setup is shown in Fig. 1. while the panel of the measurement control software is presented in Fig. 2.

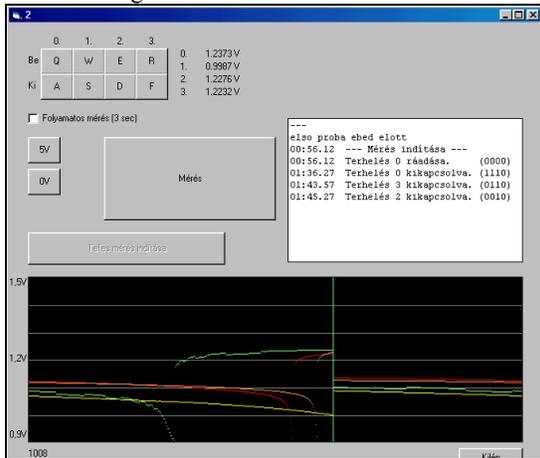

Fig. 2. Measurement control software

During the discharge period the state (the measured output voltage) is displayed for each channel and a diagram draws continuously the measured voltage-time curves.

After a discharging period, the batteries were charged with a simple four-slot, 4.8V 300mA, AA and AAA size rechargeable battery charger. This process took approximately five hours.

### 3. MODEL GENERATION

The measured results were evaluated with another computer application that was also developed for this project. This program is capable to plot the selected measurement's time-voltage curves in a prescribed range and to zoom them. The program can also calculate the effective measured capacity of the battery and display it in a chart. Time-derivative curves can also be plotted. This software is still under development in order to extend it with parts that may help to calculate the parameters of the models in the fine tuning phase.

After evaluating the measured results with this software several observations were done. First of all, we diagnosed that each measurement reproduced the Ni-MH battery's unique discharge curve shape, but the length of the period, the determining voltages and other parameters demonstrate significant differences.

A typical discharge curve consists of four different parts: (1) the initial transient, (2) the hold-period, when the voltage has got only a very small decrease during a longer period of time, (3) the exhausting transient, when voltage begins dropping and suddenly hurtles down. In this period the battery becomes fully discharged, see Fig. 3. It reaches now the preset threshold voltage very quickly, where the measurement control circuit switches off the load from the battery's terminals. (4) At this moment, the output voltage jumps to a higher value and a slow self-recharging period begins with further voltage increasing.

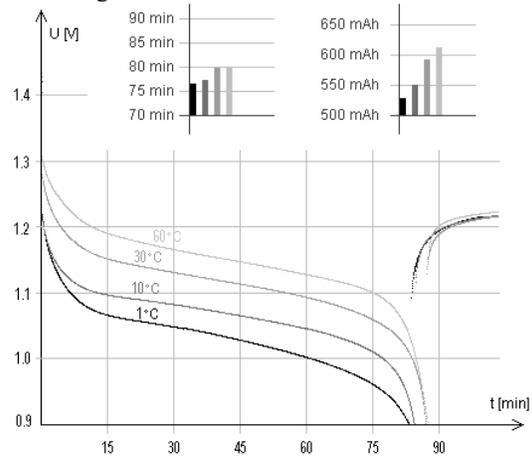

Fig. 3. Measurement results at different operating temperatures

|  | Temperature [°C] | | | |
| --- | --- | --- | --- | --- |
|  | 1 | 10 | 30 | 60 |
| Measured capacitance [mAh] | 529.2 | 552.2 | 593.6 | 614.6 |
| Discharge time [min] | 76.8 | 77.6 | 80.0 | 80.4 |

Table 1. Measurement results at different temperatures with a load resistor of 7.5 Ω.

Fig. 4. and Table 2. show, that the operating voltage and the amount of the charge drawn out from the battery

(the measured battery capacity) increase, when the load resistor is greater.

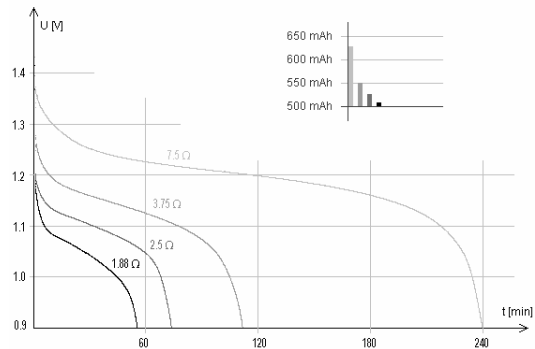

Fig. 4. Measurement results with different load resistances at 20°C.

|  | Load resistor [Ω] | | | |
| --- | --- | --- | --- | --- |
|  | 7.5 | 3.75 | 2.5 | 1.88 |
| Measured capacitance [mAh] | 630.7 | 552.6 | 528.7 | 509.6 |
| Discharge time [min] | 239,0 | 111,1 | 73,1 | 54,9 |

Table 2. Measurement results with different load resistors at 20°C.

Beyond these, the measurement results revealed a dependence of the battery's parameters, as lifetime, operating voltage, curve's slope, and resultant capacity on the time while the battery was idle between charging period finished and discharging period began. In this article we do not discuss further these parasitic effects, as in the presented measurements in order to avoid these effects we had a fixed 30 minutes idle time interval before starting the discharging process after each charging process.

Our ultimate goal is to create battery models, which will be able to provide the constant current discharge transients for arbitrary temperature conditions and load resistances. The models will be both presented by mathematical approximating functions, and the parameters of these functions will be temperature and load dependent. There are several options for the approximations.

The simplest way is the approximation with piecewise linear function [6]. The presented curves may be fitted with two-breakpoint linear functions, where the two breakpoints are the points where the original curve's curvature reaches its local maximum values. The resultant model with linear functions gives a very fast; easily to implement battery model, but the creation of it is more difficult. The curvature near the maximum point of the curve seems almost constant, and measurement inaccuracies render more difficult to determine the maximum curvature points accurately.

$$f(x) = \begin{cases} A_1 \cdot x + B_1 & x \leq x_1 \\ A_2 \cdot x + B_2 & x_1 \leq x \leq x_2 \\ A_3 \cdot x + B_3 & x_2 \leq x \end{cases}$$

Another possible way is to approach the original curve with a number of functions for the separated sections of the curve. In this case, for example, the whole curve is divided into three parts, and each part has got a specific approaching expression. The first phase of the discharge curve could be approached with a quadratic expression, the middle interval (hold-phase) with a linear slope, and the final, exhausting transient with another quadratic or cubic expression. This methodology results in more precise models, but additional problems arise. The exact position of the points, where the original curve is divided into the three separated intervals, depends on the temperature and on the load. In addition there is a problem with fitting only separated parts of the whole function: in the separator points the two neighborhood functions are joined but for optimal model performance the derivatives should be continuous as well. To fulfil all these conditions the approximating algorithms become much more complicated.

The third opportunity is to attempt with exponential convergence functions. Typical Ni-MH discharge curves can be fitted with the sum of two exponential and one linear expression, like:

$$f(x) = A \cdot e^{B \cdot x} + C \cdot e^{D \cdot x} + E \cdot x + F$$

Approaching the curves like this way, we have got the advantage of having covered the whole period with only one mathematical function, and this function approach the curve also very well. A big disadvantage of this function is that there is no exact mathematical methodology to find the fitting curve for a function, which consists of a sum of two exponential and a linear component. Most of the curve fitting algorithms, which are able to calculate the approaching exponential functions, transform the points to logarithmic scales and then fit a simple linear expression, but this methodology doesn't work for sum of two exponential functions.

An other possibility is to approach the curves with the sum of two hyperbolas and a linear component. It is similar to the previous, exponential idea, the whole curve could be covered with one expression. One hyperbola could describe the initial transient, the other is responsible for the exhausting transient, and the linear expression plays role between the two intervals, where the points are far enough from both asymptotes: in the hold-period. Sum of hyperbolas is fittable as opposed to the sum of exponential functions. The form of the function would be the following:

$$f(x) = \frac{A}{B+x} + \frac{C}{D+x} + E \cdot x + F$$

This function has got six parameters. Two parameters (*A* and *B*) determine the first parabola, two determine the second (*C* and *D*), and the remaining two form a linear function. The first parabola's right side shapes the initial transient of the battery's discharge curve (the left side is dropped). The second parabola's left curve follows the curvature of the exhausting transient, the right part of this parabola is also dropped.

After choosing the right mathematical function for the model, we can start to calculate the values of the parameters.

The hyperbolic approaching expression has got six parameters, which determines the whole function. All the six parameters of the final model (*A*, *B*, *C*, *D*, *E*, and *F*) will have the temperature and load dependence. First we take the results (the voltage–time curves) of all completed measurements, and execute the fitting algorithm on them. For each measurement we get a unique approximating function, with its own values of the six parameters. Fig. 5. shows the measured curve and the fitted hyperbolic approaching function at 30°C with 2,5Ω load, and Table 3. summarizes the resultant function parameters.

To determine the temperature and load dependent functions for the parameters, first their physical meaning should be revealed.

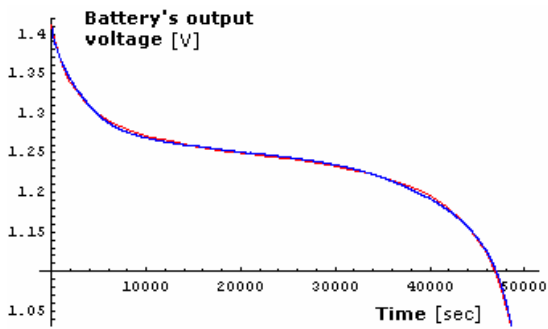

Fig. 5. Measured and fitted curve. Measurement was complete at 30°C with 2,5 Ω resistances.

| Parameters | Value |
|---|---|
| A | 230.1144 |
| B | 208.112 |
| C | 462.854 |
| D | −7310.39 |
| E | $8.39 \times 10^{-6}$ |
| F | 1.1743 |

Table 3. Parameters of the resultant function (time in seconds, voltage in Volts).

Two plus two parameters determine the two hyperbolas and two determine the linear function. In the hyperbolic expression, the parameter in the denominator defines the position of the parabolas vertical asymptote. In our application the independent variable is the time. The whole discharging period's curve starts very steeply, which means, that the approaching hyperbolas vertical asymptote is in the near. This is the *B* parameter, and that's why the value of it is always close to zero, but never equal. *B* is always a positive value, because in the *t=0* point the slope of the curve is finite, and because the asymptote is on the left from the *x* axle. The *A* parameter determines the hyperbola's curvature.

The situation is very similar at the second hyperbola. *D* parameter determines the position of the second parabolas vertical asymptote, which is close to the end of the battery's curve, where the battery became fully discharged, because here the curve's slope increases with time until the last minute. So parameter *D* is in connection with the length of the whole discharging period. Parameter *C* determines the curvature of the second hyperbola, which describes the exhausting period.

The parameter E and F describes a linear expression. Parameter *E* defines the slope of the hold-period. It is always a very small value, because in a very long period of time voltage drops only a few millivolts. Normally *E* has a negative value, as the voltage decreases with time, but sometimes it occurs, that the sum of the two hyperbolas and this line results negative gradient lineal hold-period even if *E* has got a small but positive value.

Parameter *F* is a constant, it is in physical connection the battery's operating voltage.

**Temperature and load dependence of the model parameters**

The parameter dependence on the temperature and load values shows different behavior. After evaluating the measurement results some observations could be done, as follows:

a.) the operating voltage increases with increasing the load resistance. The apposite of the battery's operating voltage is the parameter *F*. This statement states that *F(R)* function (F parameter's resistive dependence) will be a monotonously increasing function. Fig. 6. shows the results of the measurements: each point represents one F parameter of a fitted approaching function of a measured discharge curve. The diagram shows that the points have got relatively a big variance. This error comes from the nature of batteries: a certain way charged, and then same way discharged (under the same load and temperature conditions) battery can hardly reproduce the exactly the same discharge curve. It has got several reasons: the discharging behavior depends on several other factors, like how much time have spend the battery idle after is had been fully charged. Before the discharge started, was it fully discharged before the charging period started, and it also depends on the batteries past, which is obviously never can be the same if the measurements follows each other.

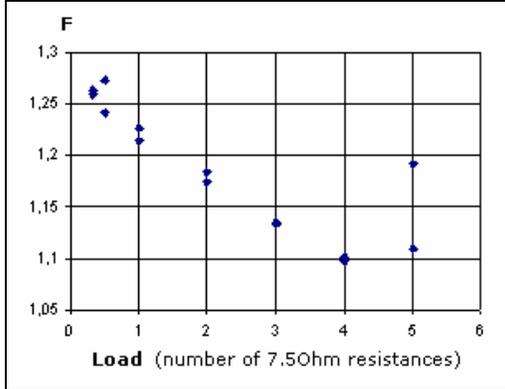

Figure 6. F parameter's dependence on load resistance values (measurement results)

b.) Another effect appears also on the diagram of Figure 6., in the range of greater resistances. Normally the points draw approximately a lineal, but in greater resistance intervals the function's gradient changes, it turns from negative to positive, and also the measurement points shows bigger variance here. This is the load range where the battery is overloaded and operating them in this range leads to significantly sorter its lifetime.

c.) When the load resistance gets greater, the length of discharging period increases. The second hyperbola's $D$ parameter is approximately in a linear connection with the length of the discharging period. If we assume, that the multiplication of the length of discharging period and the value of load resistance is constant (ideally the battery capacitance is constant), $D$ will have hyperbolic dependence of resistance. Fig. 7. shows the measured $D$ parameter points.

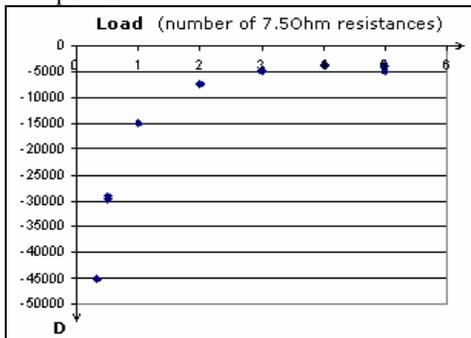

Figure 7. D parameter's dependence on load resistance values (measurement results)

Considering temperature dependence on the approaching function's parameters, further observations could be done.

d.) The operating voltage of the battery increases if we actuate it on higher temperature. The temperature dependence seems linear. $F$ parameter represents the operating voltage of the battery, that way $F$ will have linear temperature dependence. Figure 8. shows F parameter's temperature dependence.

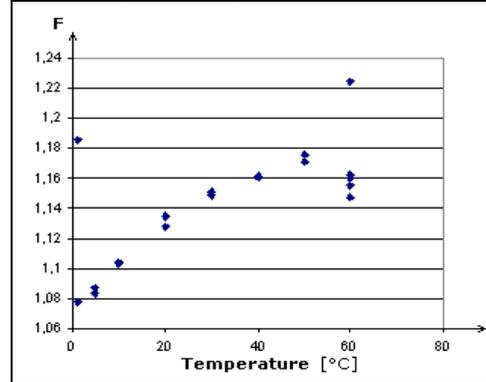

Figure 8. F parameter's dependence on temperature (measurement results)

The other parameter of the lineal component is $E$, which is the gradient value of the curves hold-period. Having a look at Figure 3. we can realize, that parameter $E$ does not has got such dependence on temperature, it will be a constant (temperature-independent) parameter.

e.) As we could see it before by the load dependent analyses, parameter $D$ in the second hyperbola represents more or less the length of the discharging period. Because the hyperbola's vertical asymptote is close to the minute when the battery became fully discharge (there is a linear correlation between these two values). The length does not show such temperature dependence, that's why parameter $D$ will also be a constant value in temperature dimension, but the curvature of the curve varies, which is represented by parameter $C$, and also with parameter $A$ by the first hyperbola. Figure 9. shows the $C$ parameters temperature dependence.

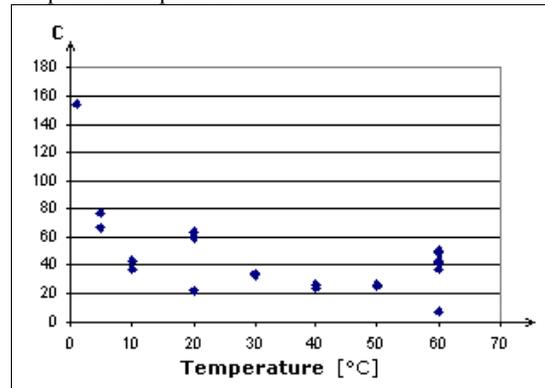

Figure 9. C parameter's dependence on temperature (measurement results)

After observing all parameters dependence on both the load and temperature dimensions, the structure of the model parameter's functions could be determined. Table 4. shows the chose functions for each parameters of the model's approaching function in temperature and load dimensions.

| Param.s | Temperature dependence | Load resistance dependence |
|---|---|---|
| A | linear | hyperbolic |
| B | linear | hyperbolic |
| C | linear | hyperbolic |
| D | linear | hyperbolic |
| E | linear | linear |
| F | linear | linear |

Table 4. Parameters of the resultant function (time in seconds, voltage in Volts).

## 4. EVALUATION OF THE MODEL

After performing the approaching procedure for each parameter, the six parameters six (two-dimensional) functions is ready.

The model we are going to prepare would be able to predict a known battery's discharge time-voltage curve under constant load and temperature conditions. These conditions are inputs of the model. The chose battery's unique parameters (which depends of its physical properties, size, capacity, and other parameters like its past) will be presented with the specific values of the parameters of *A, B, C, D, E* and *F*'s two dimensional function's.

Simulator softwares use the battery as an instrument with time, temperature and load dependence. This means that the model has three inputs: temperature, load and time, which time elapsed from the moment, when battery's discharging period has started. The only output is the battery's output voltage.

The model works with the above given approaching function. The 12 parameter's specific value is stored in a parameter table, which describes the chose battery through its properties. Substituting these twelve parameters value into the model parameter function, and also the inputs, the approaching function is able to calculate the battery's output voltage in the required time moment, under the presetted temperature and load conditions.

In the future we plan to extend our model. Two important direction will be considered: (1) observing and modeling battery's behavior under not constant (pulsing) load conditions, which is closer to the real practice, (2) creating battery models for constant performance loads (not for loads with constant resistance), as most of the mobile devices use DC/DC converters. In these batteries the output voltage drops as they are getting discharged, but the connected equipment (through the converter) has still provide the same voltage as supply voltage. So then battery output current must rise while the performance still has to stay constant. This is another aspect of the battery behavior, which is also closer to the real practice, battery model will be even more useful.

## 5. SUMMARY

Measurements with Ni-MH rechargeable batteries have been completed at different temperatures and with different constant resistive loads to examine their discharge process. The temperature was stabilized with a water basin type cooling-heating thermostat, and the measurements were controlled by a PC software, which attended to switch off the batteries load resistors when they became fully discharged. The results were evaluated with dedicated software, which was also developed by the authors.

The measurements revealed the Ni-MH battery's unique discharge curve shape. At lower temperature the Ni-MH batteries always operate on lower operation voltage, and the discharging time period becomes shorter.

The presented models are ready to be used in sensor network optimization applications.

## 9. ACKNOWLEDGMENTS

This work was supported by the PATENT IST-2002-507255 Project of the EU and by the OTKA-TS049893 and the NKFP NAP 736-205/2005 BELAMI projects of the Hungarian Government.